\documentclass[lettersize,journal]{IEEEtran}
\usepackage{amsmath,amsfonts}
\usepackage{algorithmic}
\usepackage{algorithm}
\usepackage{array}
\usepackage[caption=false,font=normalsize,labelfont=sf,textfont=sf]{subfig}
\usepackage{textcomp}
\usepackage{stfloats}
\usepackage{url}
\usepackage{xcolor}
\usepackage{verbatim}
\usepackage{graphicx}
\usepackage{cite}
\usepackage{xcolor}
\usepackage{tabularx,booktabs,textcomp}
\usepackage{amsthm}

\DeclareUnicodeCharacter{2061}{}

\usepackage{amsmath,amssymb,amsthm}

\begin{document}

\title{Data-Driven Approach to form Energy Resilient Smart Microgrids with Identification of Vulnerable Nodes in Active Electrical Distribution Network}

\author{{D. Maneesh Reddy, Divyanshi Dwivedi, Pradeep Kumar Yemula,~\IEEEmembership{Member,~IEEE}, Mayukha Pal,~\IEEEmembership{Senior Member,~IEEE}}
 

\thanks{(Corresponding author: Mayukha Pal)}
\thanks{Mr. D. Maneesh Reddy, Department of Mechanical and Aerospace Engineering, Indian Institute of Technology Hyderabad and Intern, ABB Ability Innovation Center, Hyderabad 502205, IN, (e-mail: me19btech11047@iith.ac.in).}
\thanks{Mrs. Divyanshi Dwivedi is a Data Science Research Intern at ABB Ability Innovation Center, Hyderabad 500084, India and also a Research Scholar at the Department of Electrical Engineering, Indian Institute of Technology, Hyderabad 502205, IN, (e-mail: divyanshi.dwivedi@in.abb.com).}

\thanks{Dr. Pradeep Kumar Yemula is an Assoc. Professor with the Department of Electrical Engineering, Indian Institute of Technology, Hyderabad 502205, IN, (e-mail: ypradeep@ee.iith.ac.in).}

\thanks{Dr. Mayukha Pal is a Global R\&D Leader – Data Science at ABB Ability
Innovation Center, Hyderabad-500084, IN, (e-mail: mayukha.pal@in.abb.com).}

}


\maketitle

\begin{abstract}

With the commitment to climate, globally many countries started reducing brownfield energy production and strongly opting towards green energy resources. However, the optimal allocation of distributed energy resources (DERs) in electrical distribution systems still pertains as a challenging issue to attain the maximum benefits. It happens due to the system’s complex behaviour and inappropriate integration of DERs that adversely affects the distribution grid. In this work, we propose a methodology for the optimal allocation of DERs with vulnerable node identification in active electrical distribution networks. A failure or extreme event at the vulnerable node would interrupt the power flow in the distribution network. Also, the power variation in these vulnerable nodes would significantly affect the operation of other linked nodes. Thus, these nodes are found suitable for the optimal placement of DERs. We demonstrate the proposed data-driven approach on a standard IEEE-123 bus test feeder. Initially, we partitioned the distribution system into optimal microgrids using graph theory and graph neural network (GNN) architecture. Further, using Granger causality analysis, we identified vulnerable nodes in the partitioned microgrid; suitable for DERs integration. The placement of DERs on the vulnerable nodes enhanced network reliability and resilience. Improvement in resilience is validated by computing the percolation threshold for the microgrid networks. The results show a 20.45\% improvement in the resilience of the system due to the optimal allocation of DERs.

\end{abstract}


\section{Introduction}
\label{section:Introduction}
\subsection{Motivation}

In April 2018, Puerto Rico a territory of the United States had undergone the most severe power outage incident incurred because of a fallen tree on a major power line near the town of Cayey and approximately 870,000 customers lost power supply \cite{CNN}. Such incidents are generally referred to as cascading failures; starting from a vulnerable point, significantly affecting the other parts of the electrical distribution system, and leading to power outages for many customers. {Identification of vulnerable nodes with the implementation of suitable precautionary measures could help in avoiding cascading failures \cite{vul}. Precautionary measures mainly account for the placement of distributed energy resources (DERs) at the identified vulnerable nodes that help to maintain the power supply continuity and enhance the system's resilience. The term ``Resilience" is defined as the capability of the system to withstand, respond, adapt and prevent situations such as disruptive events, man-made attacks, and severe technical faults \cite{resiliency_cost}. Generally, a resilient electrical distribution system utilizes local resources such as customer-owned solar photovoltaics (PV) and battery storage to quickly reconfigure power flows and recover electricity services during disturbance events \cite{resiliency}. However, the integration of DERs in an active electrical distribution network at vulnerable nodes would further enhance the energy resilience of the system. Thus, we referred to those vulnerable nodes as the optimal location for DERs placement for the enhancement of energy resilience. In the existing literature, the identification of these vulnerable nodes is under-evaluated in real time because of the added complexities in the electrical distribution system \cite{edison580}. In \cite{vul1} vulnerable nodes have been identified considering the uncertainty of supply and demand. In another work, vulnerable nodes have been identified in the presence of undetectable cyber-attacks in \cite{vul2}. However, the identification of the vulnerable nodes to incorporate DERs and make the electrical distribution system more resilient is not proposed.}

\subsection{Related Works}

Generally, integration of DERs is preferred at medium and low voltage sides of distribution grids because it results in minimizing the load transportation, network loss, and stress on generation and transmission systems \cite{der1}, \cite{der2}, \cite{der3}. Identification of the optimal locations for DERs integration, various methods have been proposed, and still there exists a gap in existing literature for obtaining the optimal locations with ever-increasing DERs integration and complex behaviour of electrical distribution grid \cite{der4}. Improper placement of DERs could result in voltage deviation, power quality issues, increase in power losses, harmonic distortion, and affect system reliability and resilient behaviour. Various optimization algorithms such as a multi-objective optimization planning model \cite{opt1}, a two-layer optimization method using modified African buffalo optimization \cite{african}, an ameliorated ant lion optimization \cite{dd}, an artificial hummingbird algorithm (AHA) \cite{aha}, a hybrid genetic algorithm and particle swarm optimization \cite{battery}, a data-driven optimization model for determining the location of battery energy storage system \cite{battery1} and a two-layer game model \cite{twolayer} have been used for optimal allocation of DERs in the electrical distribution system. In the state-of-art, voltage stability margin and loss reduction are only considered as key factors for the determination of the optimal location of DERs \cite{optimalsize}, \cite{optimalsize1}; and does not account for system reliability and resilience. However, enhancement in the system's resilience is an important objective to be considered for the optimal placement of DERs.

{ To measure the resilience of power distribution systems quantifiable metrics are required \cite{r1}. Thus, following the definition of resilience, various metrics have been proposed. The withstand capability of the system is given based on the aggregation of the system asset's adaptive capacity \cite{r3}. Then another metric which evaluates the impact of short-term events on long-term operational resilience is proposed \cite{r2}. Also, a probabilistic metric for quantifying the operational resilience of the electrical distribution systems for high-impact low-probability (HILP) events is proposed \cite{r8}. The above-discussed metrics only evaluate the operational resilience of the system without appending the entire infrastructure into a single quantity to provide an efficient perspective on the overall. Other features of resilience are evaluated \cite{r4} with the measures that account for the capability to anticipate, withstand, and recover for operational and planning in the electrical distribution system. Resilience metrics based on the performance of the system in terms of resourcefulness, rapidity, robustness, and adaptability are also proposed \cite{r5}, \cite{r6}.

The above-mentioned metrics correlate different characteristics of the system without considering these characteristics based on the progression of events. This leads to an improper evaluation of resilience that impacts the operation of the system. Therefore, we propose the percolation threshold as a metric that follows the progression of events and provides an effective evaluation of resilience \cite{resiliency}. The percolation threshold \cite{newman}  is a statistical measure that signifies the state transition of the system due to extreme events. It was first proposed by Flory and Stockmayer for the study of polymer condensation in the 1940s \cite{stock}. With the development of phase transition theory, scaling law and other theories, the implication of percolation theory and its critical conditions have been complemented \cite{Hammersley1964}. Percolation theory has been implemented in various fields such as chemistry, biology, geology, medicine, economics, and society to describe the dynamic behaviours of systems \cite{Blanc1986}. In the study of system reliability, the percolation theory could measure the ‘‘failure’’ degree of the system. Thus, the proposed metric is suitable for planning and operation-based energy resilience. The proposed methodology could be integrated into any digital substation as it is a data-driven analysis, on the other hand, integration of existing methods for the identification of vulnerable nodes and measuring the resilience of the system is not possible in real-world digital substations. Here, we used the Granger causality technique for the identification of the optimal locations for DERs placement in the system.}

 In early 1969, Granger discovered the most powerful and efficient statistical tool in time series analysis to find cause and effect for an econometric model before the consequence. Granger causality analyses the time series data of a signal’s predictability based on the knowledge of another signal’s history \cite{granger_2001},\cite{Granger1} using the multivariate regression model. The concept of granger causality was used in various fields including finance, neuroscience etc. An application to the field of neuroscience is when granger causality is applied to the brain's effective connectivity network to find a directional connection between brain areas or neurons \cite{brain}. The idea of causal inference based on information fusion could also help forecast electrical behaviour caused by sharing causal dependencies \cite{trans}. Considering the tool's effectiveness, we identified the network's highly dependent and correlated nodes and found them appropriate for DERs placement.

After DERs allocation to electrical distribution systems, partitioning the existing system into microgrids comes out as a second crucial step for planning and integration of DERs in traditional grids \cite{grid}. Partitioning of the electrical power system has been deliberated by many researchers using various techniques. Spectral clustering optimal partitioning technique \cite{spectral}, K-means clustering algorithm \cite{kmeans} and hierarchical agglomeration algorithm \cite{algo}. These techniques are majorly implemented in the transmission system, and due to the complex behaviour of the distribution systems, it is least explored. Here, we have utilized a graph theory-based algorithm for clustering \cite{igraph} and generating the training datasets; by identifying the clusters having maximum similarity. These obtained clustered networks are fed to Graph Neural Network (GNN) model for training to preserve the spatial correlations between the nodes of the network. It extracts similar features and clusters them in a layer-by-layer manner to facilitate the classification task. GNN is a deep neural network that operates on a graph structure and is effectively utilized for classification \cite{gnn1}. It has been extensively used in the electrical domain for the classification of fault type and their location \cite{gnnapp},\cite{gnnapp1} and for probabilistic power flow analysis in the electrical distribution system \cite{gnnapp2}.

\subsection{Contributions}

This paper presents a methodology for the optimal allocation of DERs in an electrical distribution system. This methodology could be employed in digital substation products where real-time data are readily available. For analysis and validation of the methodology, we simulated the standard IEEE-123 bus test feeder and generated the active power time series data for further evaluation. Firstly, with the help of graph theory and GNN, we have identified the optimal partitioned regions, capable to form microgrids. These partitioned regions are formed by considering similarity among the nodes to create community structures having similar nodes operating together. Further, for the obtained regions, we computed the granger causality among the nodes using time-series power flow consumption data to identify the node dependencies on each other. If dependencies are high on a particular node then its failure would lead to cascading failure in the system. These nodes are considered to be the vulnerable nodes in the system and are found suitable for the allocation of DERs to make them self-sustain. Further, we validate that the placement of DERs at vulnerable nodes enhances the system resilience and reliability by computing the percolation threshold of the network. The key contribution of this work is summarized as follows:
\begin{itemize}
    \item A novel data-driven analysis suitable for integration in the digital substation is proposed which would apprehend the real-time power flow consumption data to identify the partitions in the distribution system in terms of similarity between the nodes. This partitioning in the system lay out the self-sufficient microgrids.
    \item A novel methodology based on causal dependencies of the nodes is proposed, which helps to intuit the system's highly dependent and correlated nodes and referred it to them as vulnerable nodes. Failure of these vulnerable nodes would interrupt the power flow in the network. Thus, found them suitable for the optimal location for DERs allocation.
    \item {  Validation of the network's resilience after integration of DERs at these identified vulnerable nodes using the percolation threshold.}
\end{itemize}

This paper is organized with section \ref{section:Methods} detailing the explanation of materials and methods implemented for the proposed methodology, then section \ref{section:data} gives details about data generation and steps followed to implement the methodology. Finally, in section \ref{section:result}, the results are discussed from the implementation of the proposed methodology on the considered standard system and validated the performance of the system after execution of the proposed methodology. Section \ref{section:conclusion} concludes the study with inferences.

\section{Materials and Methods}
\label{section:Methods}
\subsection{Graph Neural Network}

Graphs are the data structures which represent the dynamic and interactive phenomena in various domains such as social networks, citation networks, chemical molecules, and recommendation systems. Graph $(G)$ comprises of edges $(E)$ and vertices $(V)$ where edges represent the relationship between the vertices.

\begin{figure}[h]
  \centering
  \includegraphics[width=3.2in]{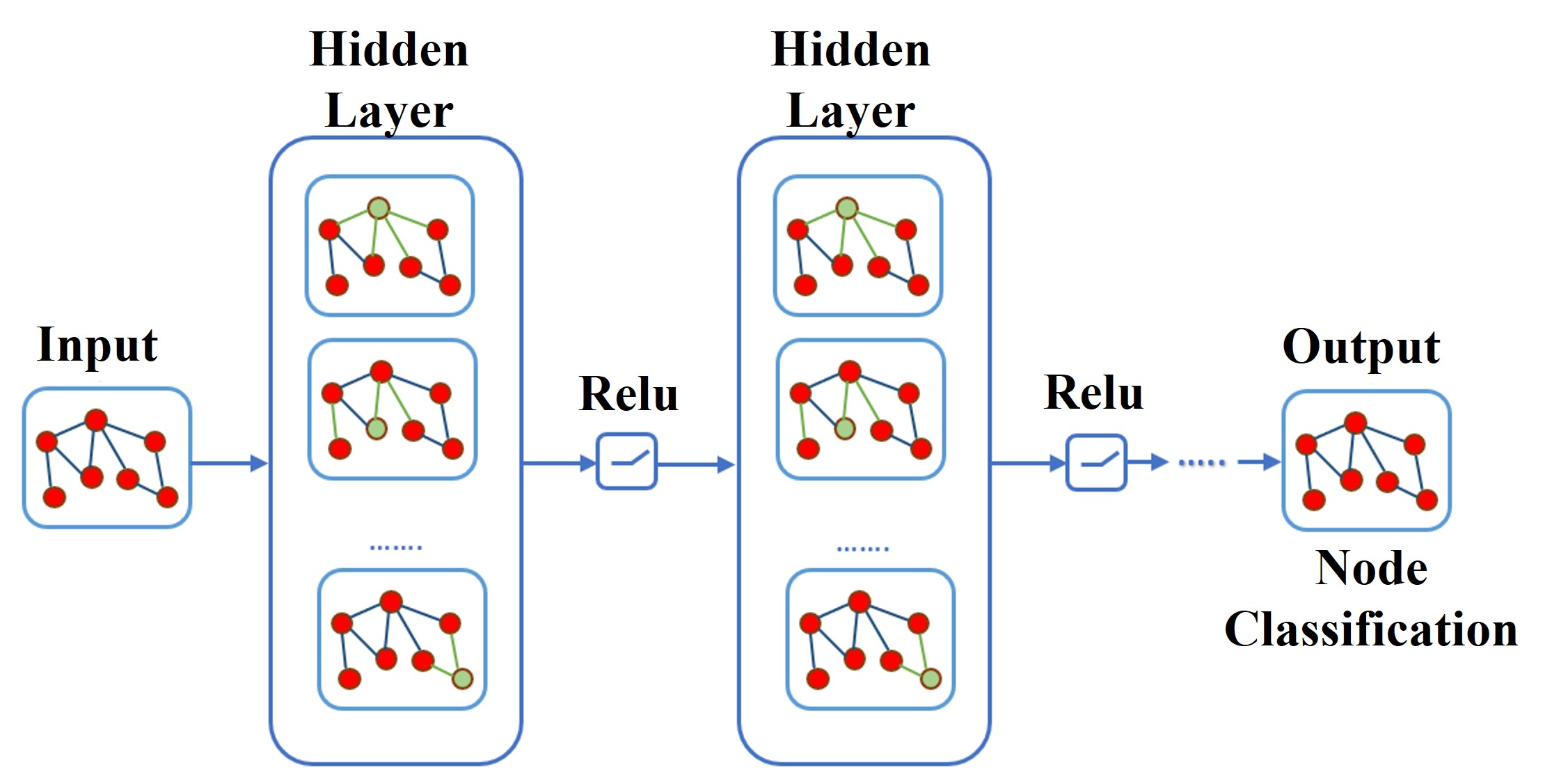}
  \caption{Process of Graph Convolution Network.}
  \label{fig:GCN}
\end{figure}

Graph Neural Network (GNN) is a type of deep learning algorithm which is implemented to process graph-structured data. This algorithm helps in node classification, link prediction, and clustering \cite{gnn1}. We have utilized the Graph Convolutional Network (GCN) for the classification of the nodes in terms of the node's similarity for community structures; the process is shown in Fig. \ref{fig:GCN}. Here, the electrical distribution system's nodes are the vertices and causal correlation (discussed in detail in section \ref{subsection:GC}) among the nodes are represented as the edges. The feature matrix corresponds to the similarity index among the nodes, and by that, the community structures will be obtained.

For a given graph $G = (V, E)$ the learning function of signals/features on a graph, GCN requires:

\begin{itemize}
    \item A $N \times N$ matrix which represents the adjacency matrix $A_m$ of the graph structure.
    \item A feature detailed for every node $i$ is given in $X_i$; which forms a feature matrix $X$ represented as $N \times M$, where $N$ is the number of nodes and $M$ is the number of input features. 
\end{itemize}

This helps in the node-level output $Z$ of size $N \times F$, where $F$ is the number of output features per node. Further, the neural network layers could be represented as non-linear functions given:
\begin{equation}
   K^{(l+1)}=f(K^{(l)},A_m)
\end{equation}
		      		
where $K^{(0)}=X$ and $K^{(l)}=Z$, $l$ is number of neural network layers. Model selection would depend on the choice of function $f(.,.)$ and its parameterization.

For this work, we consider the propagation rule as:
\begin{equation}
  f(K^{(l)},A_m)= \sigma(\hat{D}^{(-1/2)} \hat{A_m} \hat{D}^{(-1/2)} K^{(l)} W^{(l)})
\end{equation}          	     

where, $W^{(l)}$ is a weight matrix for $l^{th}$ neural network layer, $\sigma()$ is an activation function $ReLU$, $\hat{A_m} =A_m+I$, having I as the identity matrix and $\hat{D}$ is the diagonal node degree matrix of $\hat{A_m}$.

In this work, GCN layer-wise propagation rule is implemented for all nodes, $X_i \in G$, to identify the features ${h_{X_j }}$ of neighbouring nodes $X_j$. Accordingly update the  $X_i$ features using $h_{(X_j)}\leftarrow hash (\sum_j {h_{(X_j)})}$, where hash is an injective hash function. These steps are repeated until convergence. The GCN layer-wise propagation rule:
\begin{equation}
    h_{(X_i)}^{(l+1)}=\sigma \sum_j \frac{1}{c_{ij}} h_{(X_i)}^{(l)} W^{(l)}
\end{equation}
			 	       
where, $j$ indexes the neighbors of nodes $X_i$, $c_{ij}$ is a normalization constant for the edge $(X_i,X_j)$ which represents a normalized adjacency matrix. The considered parameters for the GCN network are tabulated in Table \ref{tab:parameters}. The listed hyperparameters are tuned using GridSearchCV to get the optimal values for the proposed model. The algorithm for node classification of the distribution system is given as Algorithm \ref{alg:alg1}.

\begin{algorithm}
\caption{Node classification algorithm for partitioning of electrical distribution system into microgrids using tensorflow}\label{alg:alg1}
\begin{algorithmic}
\STATE \textsc{\textbf{Input:}} Adjacency Matrix $A$, Feature Matrix $X$
\STATE \textsc{\textbf{Initialize:}} Initial\_Graph, Weight Matrix $W$\\
	     \hspace{0.5cm}train\_index, test\_index $\leftarrow$ train\_test\_split()\\    
\STATE \textsc{\textbf{while}} step=0 to Convergence do:\\
	    \hspace{0.5cm}update neural network using GCN layer\\
	    \hspace{0.5cm}model $\leftarrow$ tf.keras.models.Model(inputs=[$A$, $X$])\\
	    \hspace{0.5cm}optimizer $\leftarrow$ tf.keras.optimizers.Adam(learning\_rate)\\
\STATE \textsc{\textbf{for}} epoch in range(500):\\
        \hspace{0.5cm}with tf.GradientTape() as tape:\\
        \hspace{1cm}Compute train\_loss\\
        \hspace{1cm}grads = tape.gradient(train\_loss, weights)\\
        \hspace{1cm}optimizer.apply\_gradients(grads, weights)\\
\STATE \textsc{\textbf{if}} epoch $\%$ 50 == 0 do:\\
	    \hspace{0.5cm}mean\_loss $\leftarrow$ tf.reduce\_mean(loss)\\
        \hspace{0.5cm}accuracy $\leftarrow$ compare(test index,test label)\\
	    \hspace{0.5cm}Nodes\_classification $\leftarrow$ evaluate(all nodes)\\
\STATE \textsc{\textbf{end if}}
\STATE \textsc{\textbf{end while}}
\STATE \textsc{\textbf{Output:}}  Nodes\_classification in terms of similarity to partition the electrical distribution system into microgrids.
\end{algorithmic}
\label{alg1}
\end{algorithm}

\begin{table}
\centering
    \caption{PARAMETERS OF GCN NETWORK}
\begin{tabular}{cc}
 \toprule
Parameters & Values\\
 \midrule
Optimizers & Adam\\
Learning Rate & 0.01\\
Hidden Layers &	4\\
Layer Sizes & [4,4,2,2]\\
Epochs & 500\\
Loss & Cross Entropy\\
Activation Function & Tangent hyperbolic function\\

\bottomrule
\end{tabular}
\label{tab:parameters}
\end{table}

\subsection{Granger Causality}
\label{subsection:GC}
Granger Causality also known as ‘G-Causality’ proposes how the history of a time series variable $T_2$ aids in forecasting the future of $T_1$ such that $T_2$ ‘G-Causes’ $T_1$. This method is based on the linear formulation of a time series multivariate process. To say $T_2$ ‘G-Causes’ $T_1$ mathematically, we first construct the bivariate autoregressive (BVAR) model for the time series $T_1$ and $T_2$, where both processes must have the same time length. For analysis in an electrical distribution system, $T_1$ and $T_2$ represent the time-series real power consumption data at an interval of 1 minute. \\
$Model-1:$\\
\begin{equation}
    T_1 (t)=\sum_{(l=1)}^n{\alpha_{(11,l)} T_1 (t-l)} +\sum_{(l=1)}^n {\alpha_{(12,l)} T_2 (t-l)}
    +\varepsilon_1 (t)
\end{equation}

\begin{equation}
        T_2 (t)=\sum_{(l=1)}^n{\alpha_{(21,l)} T_1 (t-l)} +\sum_{(l=1)}^n {\alpha_{(22,l)} T_2 (t-l)}+\varepsilon_1 (t)
\end{equation}

where $n$ is the model order or the maximum number of lags that is incorporated in our model which can be found through computing the Akaike Information Criterion (AIC) \cite{aki} and Bayesian Information Criterion (BIC) \cite{bsi}. The coefficients $\alpha_{(ij,l)}$ $(i,j=1,2$ and $l = 1,2,...,n)$ also known as model parameters can be found by Yule-Walker equations. $\varepsilon_1 (t)$ and $\varepsilon_2 (t)$ are the error terms accounts for loss of variance.
Then, we construct the univariate autoregressive model for the time series process $T_1$ and $T_2$. By removing $T_2$ in equation (4) and $T_1$ in equation (5), we could obtain Model 2.\\
$Model-2:\\$
\begin{equation}
    T_1 (t)=\sum_{(l=1)}^n{\alpha_{(11,l)}^*} A_1 (t-l)+\varepsilon_{11}(t)
\end{equation}
 
\begin{equation}
    T_2 (t)=\sum_{(l=1)}^n{\alpha_{(22,l)}^*} A_2 (t-l)+\varepsilon_{22}(t)
\end{equation}

If the error terms $\varepsilon_{11},\varepsilon_{22}$  are increased in comparison to error terms, $\varepsilon_{1},\varepsilon_{2}$ then we can conclude that $T_2$ ‘G-causes’ $T_1$  and $T_1$ ‘G-causes’ $T_2$  respectively. From the logarithmic ratios of the error terms, one could find the magnitude of ‘G-causality’. For illustration purposes, if $T_2$ ‘G-causes’ $T_1$ then the magnitude of ‘G-causality’ is given by:

\begin{equation}
    \mathcal{F}_{2 \rightarrow 1}=log \frac{var(\varepsilon_{11}(t))}{var(\varepsilon_{1} (t))}⁡
\end{equation}

\begin{figure*}
  \centering
  \includegraphics[width=5.6in]{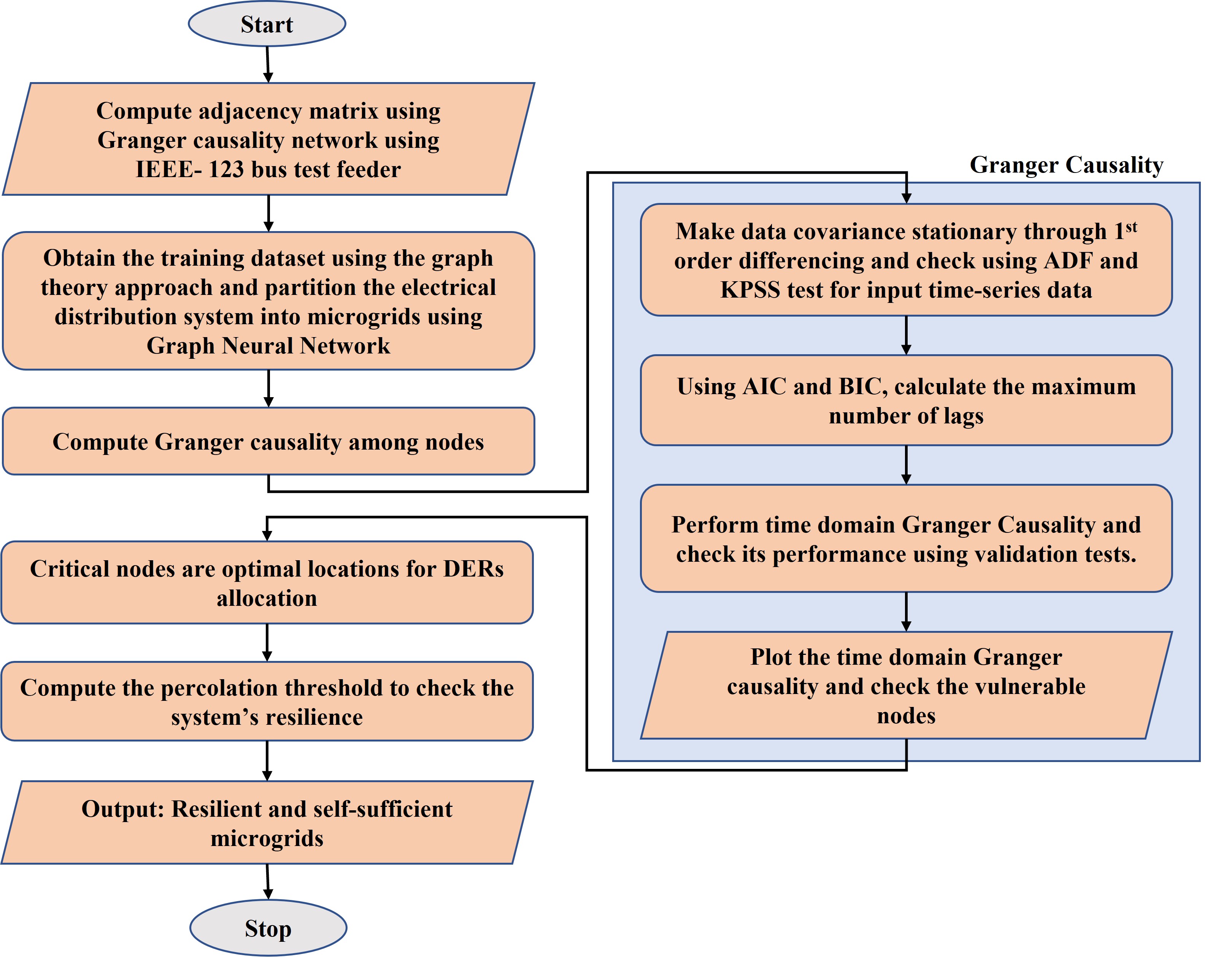}
  \caption{Proposed Methodology with the steps of the implementation of Granger Causality, identification of vulnerable nodes and evaluation of system resilience.}
  \label{fig:FC2}
\end{figure*}

If there are more time series processes involved (i.e., $T1$, $T2$, $T3$, $T4$ and so on) then we use the same approach as discussed above where instead of BVAR we use multivariate autoregressive (MVAR) modelling which is also referred as conditional G-Causality \cite{Granger1}.
One primary assumption associated with all the modelling is that the time series must be covariance stationary. Covariance stationery is a condition where the time series have the same mean, variance, and the covariance between any two of them is exclusively determined by their relative locations i.e., how far apart they are from one another. A step-by-step implementation of the Granger causality in our proposed methodology is shown in Fig. \ref{fig:FC2}. With the implementation of Granger causality, we would identify the most causing nodes in the partitioned sections of a distribution network system and suggest those nodes suitable for DERs placement.

\subsection{Percolation and System resilience}
\label{subsection:PT}
Further, for validating the resilience of the electrical system when DERs are allocated at the identified locations, the percolation threshold is computed as a measure of resilience \cite{resiliency}. The percolation threshold is an important computational tool of network science that helps to identify the transitions in the system’s operating conditions. Percolation theory is found to be effective for obtaining both quantitative and qualitative measures of the resilience of networks \cite{resiliency}. Computation of the percolation threshold is achieved by calculating the percolation strength of the network, followed by calculating susceptibility and the best estimate of the percolation threshold $\rho_c$ as the value of $\rho$ is where the susceptibility reaches its maximum \cite{dd}. 

\noindent Percolation Strength,
\begin {equation}
    P_\infty (p) = \frac{1}{NQ} \cdot \sum_{q=1}^{Q} S_q(p)
\end {equation}

where, $S(p)$ as a function of bond occupation probability, $p=e/E$, with $E$ the total number of edges and $e$ is the number of edges removed from the initial configuration. $N$ is the number of nodes. Then, the Susceptibility

\begin {equation}
 	\chi(p)=  \frac{(1/N^2 Q)\sum_{q=1}^{Q} (S_q(p))^2-[P_\infty (p)]^2} {P_\infty (p)} 	 
\end {equation}
 
\begin {equation}
    \rho_c=arg{[\max \chi(p)]}
\end {equation}

Thus, we first created complex networks with and without the integration of DERs. Then using equation (9), computed the bond occupation probability of the network by removal of network’s edges/vertices and calculated the percolation strength of the network. Finally identified the bond occupation probability where the susceptibility is maximum and refer to it as the percolation threshold as given in equation (11).

This percolation threshold is the quantifiable measure of resilience and for an electrical system, a high percolation threshold value refers more resilient system. Thus, this methodology is utilized for validation of the proposed framework for identifying the optimal location for DERs with ensured system resilience.

\section{DATA AND ITS PROCESSING}
\label{section:data}
In this section, we discuss the used simulated data and the proposed methodology followed for the framework. Firstly, we consider a standard IEEE-123 bus test feeder that operates at 4.16kV and modelled it using an electrical distribution system's simulation platform known as GridLAB-D. This simulation tool helps in generating the time-series data for our desired system with required DERs, load etc. IEEE-123 Node Test Feeder consists of 85 constant loads, we collected incoming real power on 36 nodes and these nodes are referred to as Meter nodes in our further analysis.

\begin{figure}
  \centering
  \includegraphics[width=3.3in]{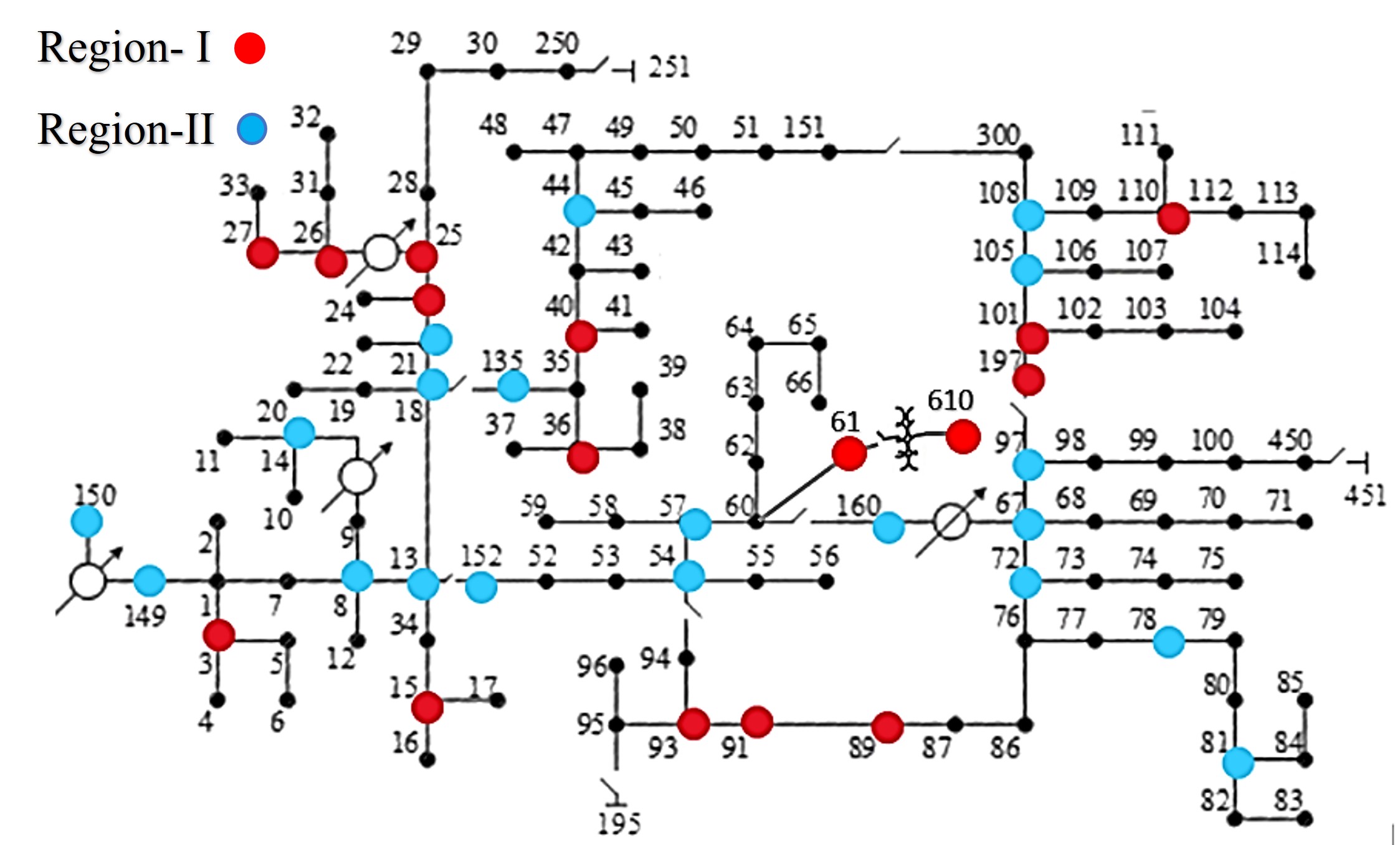}
  \caption{IEEE-123 bus test feeder, optimally clustered as Region-I and Region-II.}
  \label{fig:IEEE}
\end{figure}

From the generated time-series real power load consumption data, Granger causal network was created. Then, the network was analyzed using optimal cluster based on the graph theory and GNN to obtain the two optimally clustered microgrids in terms of similarity in the IEEE-123 bus test feeder; referred to them as Region-I and Region-II as shown in Fig \ref{fig:IEEE}. 

Furthermore, to identify the optimal location for installing DERs in the system, we performed the Granger causality to identify the functional dependencies among the nodes of the regions. Nodes having highly functional dependence would significantly impact the other nodes of the region. This would affect the reliability of the network, as an interruption in those nodes would lead to system outages or cascading failures. Thus, the identification of highly dependent nodes of the network and then incorporating DERs (mainly solar PV panels of $50kW$ each) at those nodes would enhance the reliability and resilience of the electrical distribution system.

Finally, for validating the identified optimal locations for DERs incorporation for system resilience, we computed the percolation thresholds for both regions, with and without the incorporation of DERs. Firstly, we calculated the Pearson correlation coefficient among the nodes, then considered positive coefficient thresholding to obtain the adjacency matrix which helps in forming the complex network. The obtained complex networks are used to compute the percolation thresholds of the network as detailed in section \ref{subsection:PT}. The proposed methodology is shown in Fig. \ref{fig:FC2} and further, the obtained results were discussed in the next section.

\section{Results and Discussion}
\label{section:result}

With the help of GNN, we computed the optimal community clustered structures for the IEEE-123 bus test feeder as shown in Fig. \ref{fig:IEEE}. The classification accuracy for the partitioned clusters as in Fig. \ref{fig:IEEE} is 86\%, which is the highest in comparison to other possible clustered structures. Thus, the network was efficiently partitioned into two regions, when DERs are incorporated in both regions then the system would operate as two optimally partitioned microgrids. Furthermore, to identify the optimal allocation of DERs in these two regions, granger causality is used. We also validated the network resilience and reliability with incorporated DERs at identified vulnerable nodes using a percolation threshold. A complete end-to-end data-driven system architecture is designed using the proposed methodology for a resilient, reliable microgrid with optimal DERs allocation at identified vulnerable nodes in the network.

\subsection{Analysis of the partitioned Region-I}
\begin{figure}[h]
  \centering
  \includegraphics[width=2.2in]{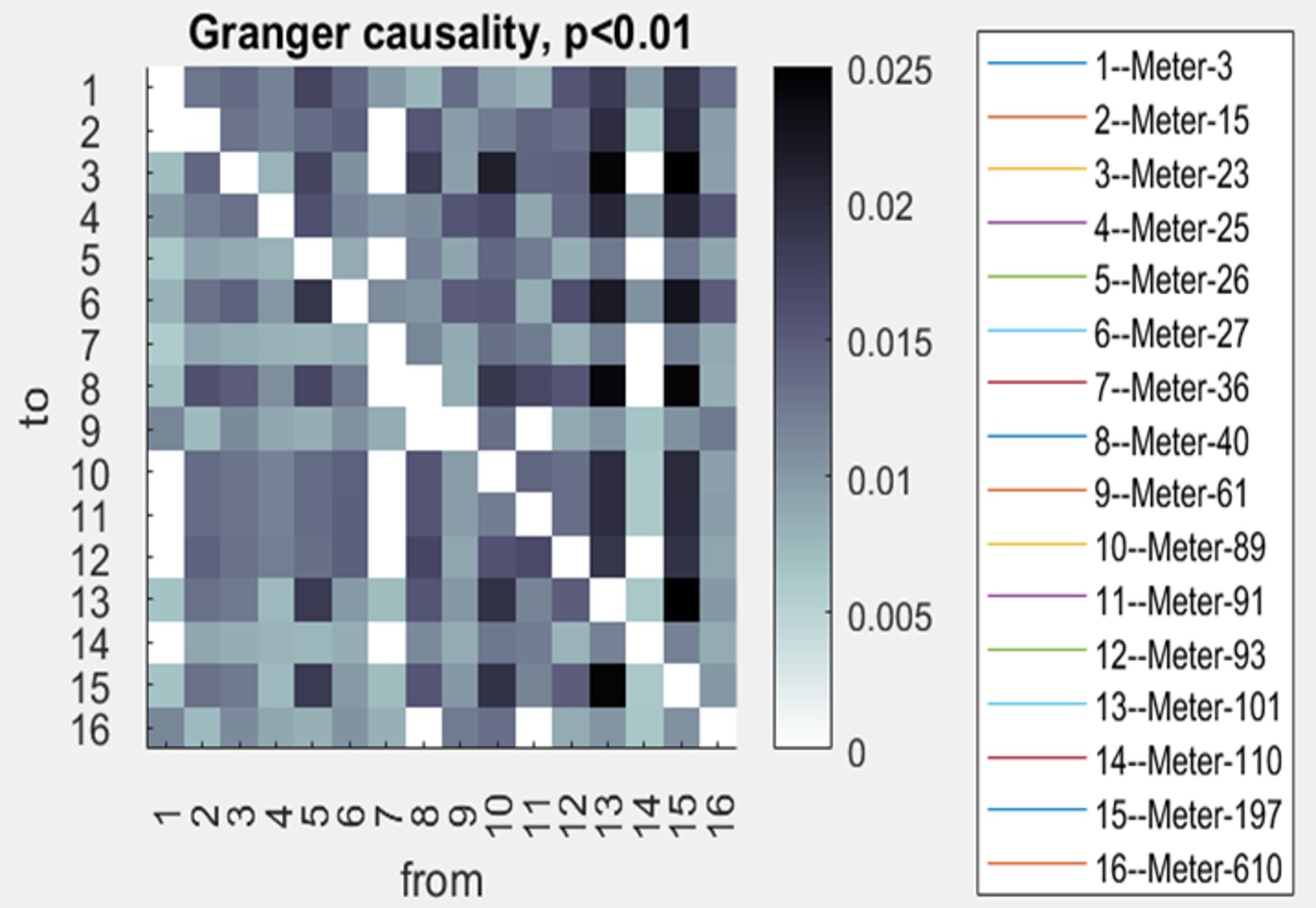}
  \caption{Granger causality analysis for Region-I; Meter-101, Meter-197, Meter-26, and Meter-89 are the nodes which are dominantly causing all other nodes of the region-I.} 
  \label{fig:R1_without}
\end{figure}

Initially, Granger causal analysis was performed on Region-I without the incorporation of any DERs in the system. From Fig. \ref{fig:R1_without}, we can identify that Region-I nodes are strongly connected and are of very high magnitude. It signifies that if any node fails then it would affect the other nodes of the system and lead to interruption in the supply or system failure. For Region-1, among all 16 nodes of this region, we saw four nodes named Meter-101, Meter-197, Meter-26 and Meter-89 are strongly causing all other nodes of the region; identified them as vulnerable nodes of this region. Hence,  incorporated DERs like solar panels at these nodes.

\begin{figure}[h]
  \centering
  \includegraphics[width=3.3in]{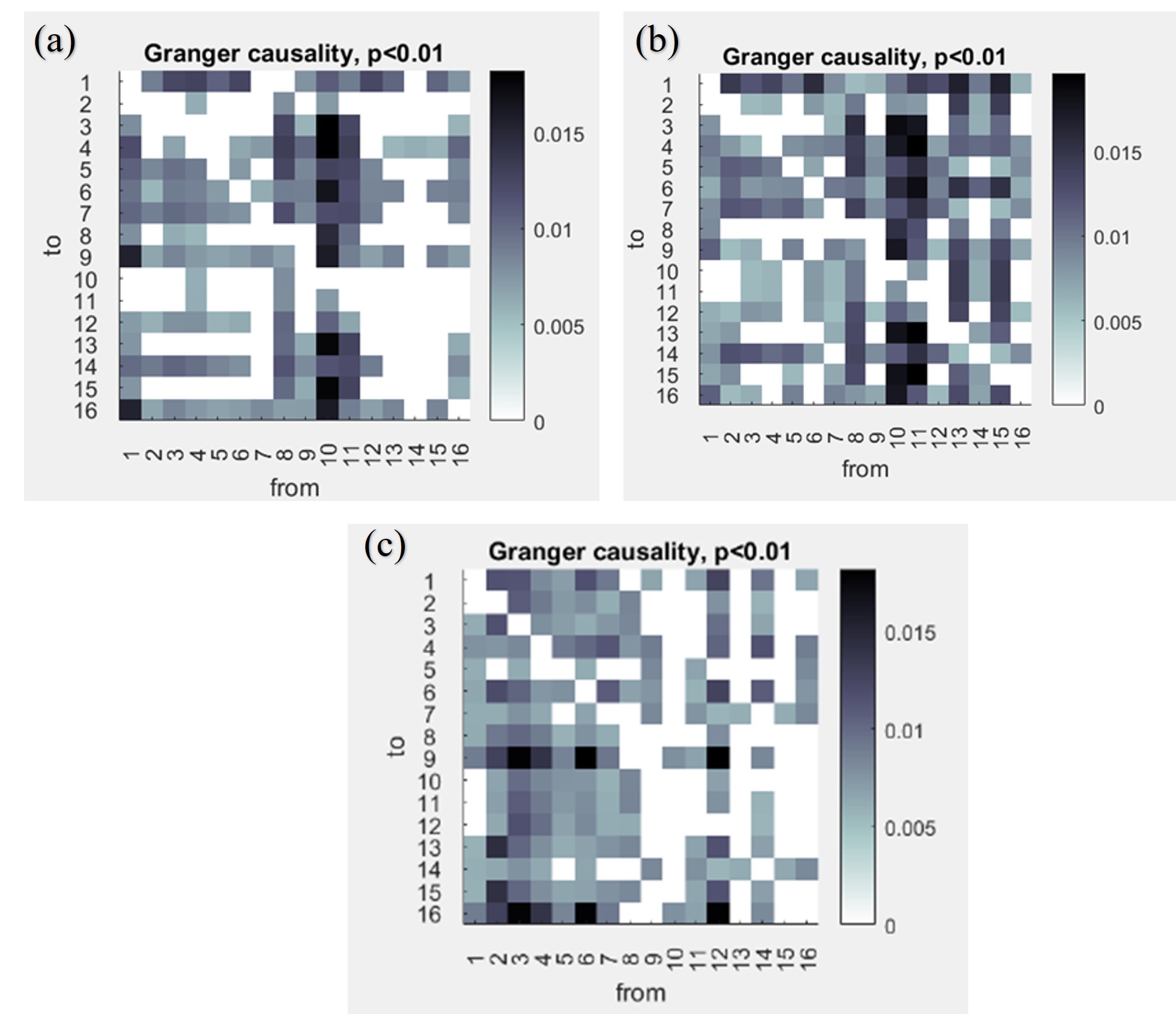}
  \caption{Granger causality analysis with solar PV panels incorporation in Region-I at (a) Meter-101, where the significant drop in causal dependencies is observed, (b) Meter-197, dependencies among vulnerable nodes persist, and (c) all vulnerable nodes including Meter-101, Meter-197, Meter-26, and Meter-89, magnitude for causal dependencies among the nodes got dropped significantly.}
  \label{fig:R1_with}
\end{figure}

Further, we incorporated solar PV panels only at Meter-101 and from Fig. \ref{fig:R1_with}(a), we observe a significant drop in causing effect among the nodes along with the causing magnitude which suggests that the system nodes now became less dependent on other nodes. In case of any failure, other nodes can sustain and operate without being intervened because of other nodes. Similarly, we incorporated solar PV panels at Meter-197 and from Fig. \ref{fig:R1_with}(b), we observe that causing magnitude is dropped significantly and the dependency also compared to the region operating without DERs.

Finally, we analyzed the system response with the incorporation of solar PV panels at all these three vulnerable nodes of the system. From Fig. \ref{fig:R1_with}(c), we observe that the overall network dependencies were reduced. With the incorporation of solar PV panels at the vulnerable nodes, the microgrid’s nodes became more capable to maintain the supply-demand without a requisition from other nodes. Thus, the proposed methodology for optimal allocation of DERs in Region-I enacted the system to be more reliable, self-sufficient, and capable to operate as a microgrid.

\subsection{Analysis for the partitioned Region-II}
\begin{figure}[h]
  \centering
  \includegraphics[width=2.2in]{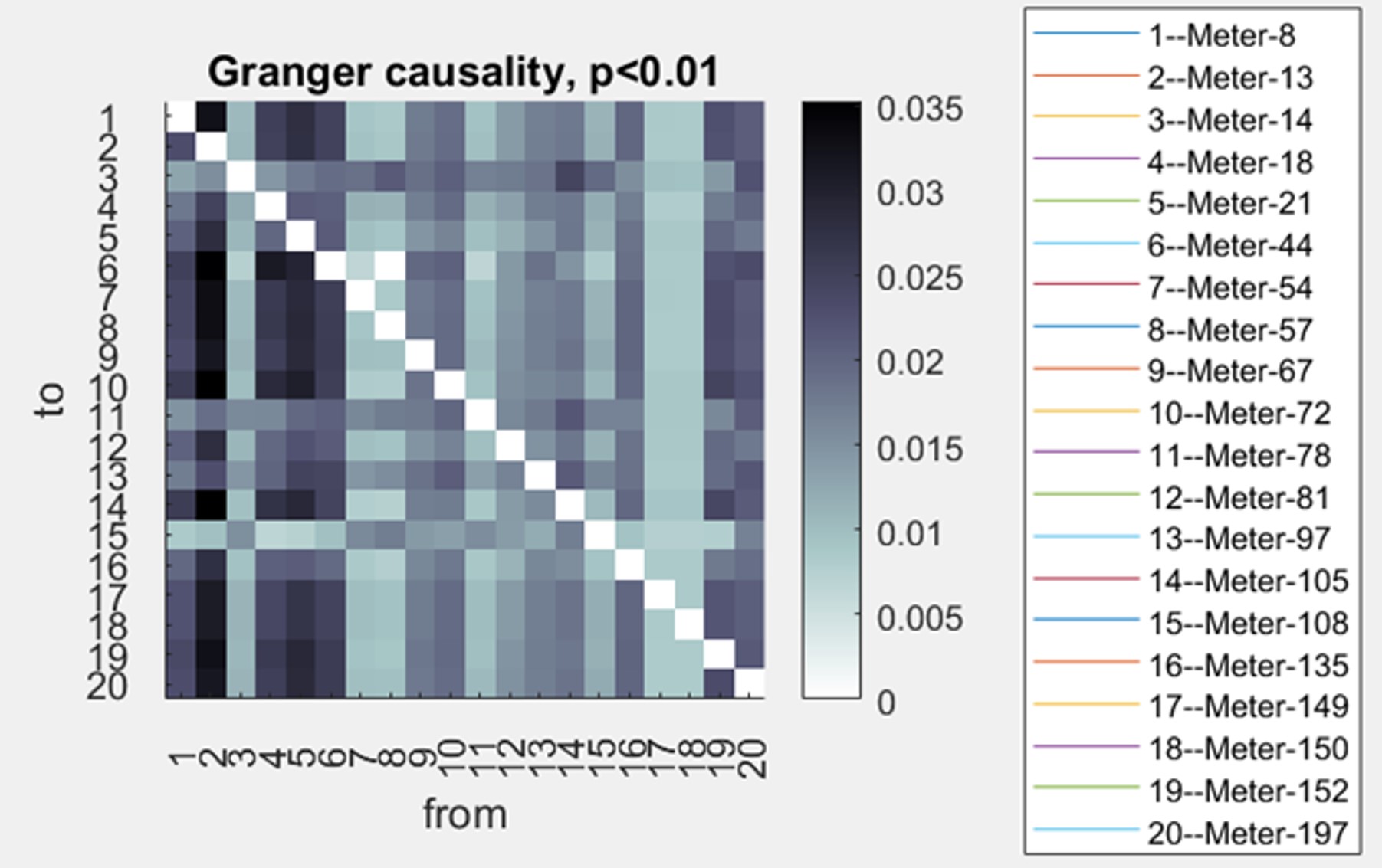}
  \caption{Granger causality analysis for Region-II; Meter-13, Meter-21, Meter-18, and Meter-44 are the nodes which are dominantly causing all other nodes for Region-II.}
  \label{fig:R2_without}
\end{figure}

A similar analysis for Region-II is also performed. Fig. \ref{fig:R2_without}, which shows the causal dependencies among the nodes without incorporation of DERs and we observe that the causing magnitudes among nodes are high. Here, Meter-13, Meter-21, Meter-18, and Meter-44 are identified as the vulnerable nodes considered for incorporating solar PV panels for manifesting reliability in the region.

\begin{figure}[h]
  \centering
  \includegraphics[width=3.3in]{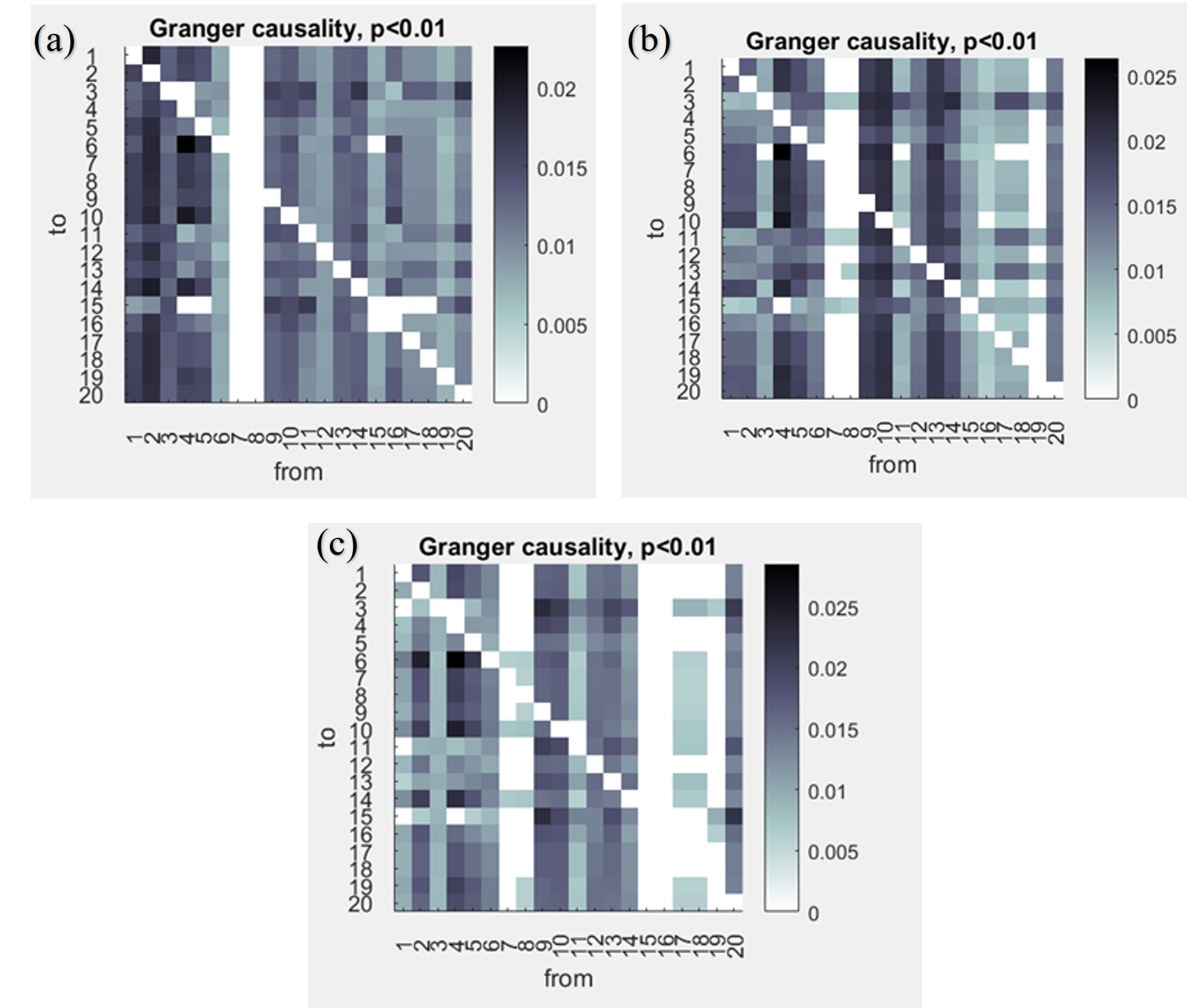}
  \caption{Granger causality analysis with solar PV panels incorporation in Region-II at (a) Meter-21, where the significant drop in causing magnitude and causal dependencies from Meter-54 and Meter-57, (b) Meter-13, drop in causal dependencies from Meter-54, Meter-57, and Meter-152, and (c) all vulnerable nodes including Meter-13, Meter-21, Meter-18, and Meter-44, magnitude and causal dependencies among the nodes got dropped significantly.}
  \label{fig:R2_with}
\end{figure}

With the incorporation of solar PV panels only at Meter-21, from Fig. \ref{fig:R2_with}(a) we identified that causing magnitude significantly dropped and Meter-54 and Meter-57 become independent and does not have any causing impact on other nodes. Then with the incorporation of solar PV panels only at Meter-13. From Fig. \ref{fig:R2_with}(b) we can observe that the magnitude is slightly more as compared to Fig. \ref{fig:R2_with}(a) but dependency among nodes got reduced. Finally, incorporated solar PV panels at all the vulnerable nodes. From Fig. \ref{fig:R2_with}(c) it is observed that causing magnitude is significantly lowered as compared to even Fig. \ref{fig:R2_without} scenario and more nodes became independent. 

Thus, with the incorporation of solar PV panels, Region-II results made it a more reliable distribution microgrid which is self-sufficient and could be able to maintain supply-demand efficiently.

\begin{table}[h]
\centering
    \caption{COMPUTING PERCOLATION THRESHOLD FOR REGION-I AND REGION-II WITHOUT AND WITH DERs INTEGRATION}
\begin{tabular}{ccc}
 \toprule
Regions	& DERs Incorporation & Percolation Threshold\\

 \midrule
Region-I &\hspace{0.15cm}Without DERs &	0.06722\\
         &\hspace{0.20cm}With DERs at Random Nodes	& 0.07733\\
	   &\hspace{0.20cm}With DERs at Vulnerable Nodes	& 0.09766 \vspace{0.25cm}\\   
 
Region-II &\hspace{0.15cm}Without DERs & 0.05263\\
        &\hspace{0.20cm}With DERs at Random Nodes	& 0.05868\\
	&\hspace{0.20cm}With DERs at Vulnerable Nodes &	0.08349\\
\bottomrule
\end{tabular}
\label{tab:percolation}
\end{table}

\subsection{Validation of the proposed methodology}

In this section, we have validated the reliability and resilience of the system when DERs got incorporated at the vulnerable nodes. This validation is implemented by computing the percolation threshold for the correlated networks of regions without and with DERs. The percolation threshold helps in identifying the resilience of the networks. For an electrical system, if the percolation threshold is high then we infer that the system is more resilient. We have computed the percolation threshold for Region-I and Region-II. Table \ref{tab:percolation} shows that the percolation threshold value increased when DERs incorporated at the vulnerable nodes in both regions. We have also incorporated the same number of solar PV panels of the same rating in both regions randomly and observed that the improvement in percolation threshold is more significant when DERs are placed at vulnerable nodes.

Thus, this observation proves the significance of the proposed methodology for identifying the optimal allocation of DERs. With DER incorporation at those locations in the network, the resilience was enhanced, and the system attains the ability to sustain unwanted interruption without breaking down the entire system.

\section{Conclusion}
\label{section:conclusion}

We proposed an effective methodology for the optimal allocation of DERs in the electrical distribution grid with the system's resilience quantification and validation. This optimal location for DERs is identified in terms of vulnerable nodes of the electrical distribution system which are highly vulnerable and subjected to any disturbances that lead to outages or failure. Initially, we partitioned the standard IEEE-123 test feeder into optimal clusters using graph theory, created training datasets and used a graph neural network to optimally partition the system considering node similarity as a feature to form community structure. Thus, we clustered the network into two regions such that when DERs incorporated into the system the clustered regions would act as self-sustained microgrids. 

Further, we identified the vulnerable nodes of two regions using Granger causality analysis, which helped in identifying the causal dependencies among the nodes of the regions. The strongly causing nodes of the system are considered the vulnerable nodes of the regions and thus get easily affected by the variation in other nodes of the region. These identified vulnerable nodes are the optimal locations for incorporating the DERs in the system as they are highly sensitive to variation due to other nodes in the system. Then, we incorporated solar PV panels at these vulnerable nodes and again computed the granger causality for the region to observe that the casual dependencies and causing magnitudes among nodes are significantly reduced for both the partitioned regions.

Finally, we validated the results obtained by Granger causality, proving that the placement of DERs at identified vulnerable nodes of the regions resulted in effective microgrids. This validation is achieved by calculating the percolation threshold of both the regions without and with DERs. The obtained result proves the effectiveness of the proposed methodology as percolation threshold values increased by 31.17\% for Region-I and 36.96\% for Region-II when the system is incorporated with DERs at vulnerable nodes. Thus, it is proven that the identified optimal location for placing the DERs results in reliable and resilient microgrids. Also, the node dependency is reduced significantly thus the system nodes are self-sufficient and capable of maintaining the supply-demand independently.

In future, we plan to extend this work for identifying the optimal size and type of DERs allocation in electrical distribution systems using data-driven approaches.

\bibliographystyle{IEEEtran}
\bibliography{main}

\end{document}